\documentclass[a4paper, amsfonts, amssymb, amsmath, reprint, showkeys, nofootinbib, twoside, superscriptaddress]{revtex4-1}
\usepackage[english]{babel}
\usepackage[utf8]{inputenc}
\usepackage{xcolor}
\usepackage{graphicx}
\usepackage[colorlinks, pdftex, pdftitle={Article}, linkcolor=blue, pdfauthor={Author}]{hyperref} 
\usepackage{longtable}
\usepackage{appendix}
\usepackage{tabularx}
\usepackage{ltxtable}
\usepackage[caption=false]{subfig}

\begin{document}

\newcommand\todo[1]{\textcolor{red}{#1}}
\newcommand\bryan[1]{\textcolor{orange}{#1}}
\newcommand\olivia[1]{\textcolor{violet}{#1}}
\newcommand{\tket}{t$\mid$ket$\rangle$ }
\newcommand{\tkets}{t$\mid$ket$\rangle$}

\newcommand{\unit}{\mathds{1}}

\def\sectionautorefname{Section}
\def\subsectionautorefname{Subsection}
\def\appendixautorefname{Appendix}

\title{A QUBO formulation for qubit allocation}

\author{Bryan Dury}
    \email[Correspondence email address: ]{b.dury@alumni.ubc.ca}
    \affiliation{Department of Physics and Astronomy, University of British Columbia, Vancouver, Canada}
    \affiliation{TRIUMF, Vancouver, Canada}
    
\author{Olivia Di Matteo}
    \email[Correspondence email address: ]{odimatteo@triumf.ca}
    \affiliation{TRIUMF, Vancouver, Canada}

\date{\today}

\begin{abstract}
To run an algorithm on a quantum computer, one must choose an assignment from logical qubits in a circuit to physical qubits on quantum hardware. 
This task of initial qubit placement, or \emph{qubit allocation}, is especially important on present-day quantum computers which have a limited number of qubits, connectivity constraints, and varying gate fidelities. 
In this work we formulate and implement the qubit placement problem as a quadratic, unconstrained binary optimization (QUBO) problem and solve it using simulated annealing to obtain a spectrum of initial placements.
Compared to contemporary allocation methods available in \tket and Qiskit, the QUBO method yields allocations with improved circuit depth for $>$50\% of a large set of benchmark circuits, with many also requiring fewer CX gates. 
\end{abstract}

\maketitle

\section{Introduction}

The past decade has seen significant development in quantum computing hardware, with a number of commercially-available machines and software libraries that enable users to program and execute their own quantum algorithms. 
While architectures and implementations vary, common issues with present-day machines are the limited qubit connectivity and high error rates, especially for two-qubit operations.

A crucial underlying part of the quantum software stack is the process of quantum compilation, which includes circuit synthesis and optimization, transpilation, initial placement, and qubit routing.
Initial placement, or \emph{qubit allocation}, is the process that assigns logical qubits in a quantum circuit to physical qubits on the quantum hardware graph. 
This must be done taking into account a variety of factors: error rates, number of operations, operation times, decoherence times, and connectivity all play a role in determining the quality and success of a quantum algorithm.

The problem of qubit allocation, however, is NP-complete \cite{Siraichi2018}. 
While for small cases we can simply test every possible allocation and determine the one with, e.g. the highest success probability, or the fewest SWAPs to work around connectivity, for larger circuits and devices one must design effective techniques to choose the allocation. 
This problem has garnered significant attention lately, with a variety of approaches considered \cite{Tan2020-2, Dueck2018, Li2020, Ghosh2020, Zhang2020, Zhang2020-2, Childs2019, Zulehner2019, Brierley2015, Paler2018, Paler2020, Nash2019, Lin2019, Smith2020a, Webber2020} and incorporated in a number of full-stack toolkits \cite{Amy2020, Smith2020, Sivarajah2020, Qiskit, cirq}.
Some common techniques involve partitioning of the circuit into blocks, finding the optimal assignment within each block, and then swapping qubits between blocks to satisfy the connectivity constraints \cite{Wille2019, Li2018, Cowtan2019, Pedram2016}.
Other approaches use machine-learning techniques to optimize circuit synthesis \cite{Herbert2018a, Pozzi2020}.
Recently, a number of methods have focused on incorporating hardware calibration data in an effort to improve the final circuit fidelities \cite{Murali2019, Wilson2020, Jurcevic2020, Nishio2019, Tannu2018, Bhattacharjee2019} .

A handful of approaches have also considered simulating annealing \cite{Finigan2018, Zhou2020}.
Simulated annealing is a widely-applied heuristic optimization technique that involves randomly choosing an initial configuration, and allowing the system to transition between states with some probability in order to find a global minimum. 
In this work, we apply simulated annealing to the qubit allocation problem formulated specifically as a quadratic, binary unconstrained optimization problem, or QUBO.

The QUBO formulation is familiar to the quantum computing community due to its equivalence to the two-dimensional Ising model, and its use as the basis for the optimization problems solvable by D-Wave's quantum annealers and Fujitsu's digital annealers. A cost function for a QUBO problem can be expressed in the form:
\begin{equation}
    \min_{x} \sum_{ij} \sum_{k\ell} Q_{ijk\ell} x_{ij} x_{k\ell} + \sum_{ij} b_{ij} x_{ij} + \hbox{constraints},
\end{equation}
where $x_{ij}$ are binary variables for which we would like to find an assignment, and $Q_{ijkl}$ and $b_{ij}$ are coefficients that incorporate information about relationships between them. The aim is to find an assignment of $x_{ij}$ such that the above cost, or `energy', is minimized.
For qubit allocation we use the $x_{ij}$ to indicate the decision to assign logical qubit $i$ to hardware qubit $j$ ($x_{ij} =1$ if true, $x_{ij} = 0$ if not). 
The quadratic terms $Q_{ijkl}$ carry information about the quality of choosing, in the same assignment, to map logical qubit $i$ to hardware qubit $j$, and $k$ to $\ell$.
Similarly, the linear terms are based on the quantity and quality of single-qubit operations. Constraints are added to ensure unique assignment of each logical qubit to a single hardware qubit.

The QUBO method coupled with simulated annealing demonstrated a number of distinct advantages compared to existing initial allocation methods.
The method was found to be incredibly flexible in terms of cost-function design.
For example, including a specific metric (e.g. success probability) within the QUBO cost function results in higher quality allocations with respect to that metric.
Simulated annealing also enables one to generate thousands of solutions for relatively low computational cost.
Having access to these distributions of initial allocations allowed us to investigate the allocation process in more detail, as we could look at where logical qubits tended to be allocated on the hardware graph.
Finally, the computational requirements are agnostic to the structure of the circuit, and essentially limited by only the time-scaling of simulated annealing.
The largest circuits we tested were on a 53-qubit hardware graph, with 1000 allocations obtained in 30 minutes (2 seconds per allocation), making this method appealing for the next generations of NISQ devices that will have in the range of 50-100 qubits. 

In \autoref{sec:formalism}, we discuss QUBOs, our choice of coefficients, and the metrics by which we gauge the quality of our solutions. 
In \autoref{sec:implementation} we analyze the performance of our method using a benchmark circuit set, and provide details about our implementation, which we note is available open-source on our Github \cite{github}. 
\autoref{sec:benchmarks} compares the solution quality of the QUBO formulation to that of other contemporary software tools --- Qiskit \cite{Qiskit} and \tket \cite{Sivarajah2020} --- as well as  the recently published QUEKO benchmarks \cite{Tan2020} to investigate the optimality gap of QUBO allocations.
We conclude in \autoref{sec:conclusion} by suggesting a number of interesting possible extensions of this method.


\section{A QUBO for qubit allocation}
\label{sec:formalism}

\subsection{QUBO formalism}
A quadratic unconstrained binary optimization problem, or QUBO, is generally defined as
\begin{equation}
    \min_{x} \enskip x^{T} Q x
    \label{eq:qdef}
\end{equation}
where $x$ is an $N$-dimensional vector of binary variables and $Q \in \mathbb{R}^{N \times N}$. 
The QUBO model can represent a variety of problems in the field of combinatorial optimization (CO) such as max-cut, set partitioning, graph-colouring, and quadratic assignment \cite{Kochenberger2014}.
An excellent tutorial paper on the formulation of QUBO models for problems like these is available in \cite{Glover2018}.
In particular, the QUBO model is equivalent to the Ising model up to a linear transformation, allowing many problems in the physics domain to be recast in it as well \cite{Lucas2014, Lodewijks2019}.

A particular strength is that once a problem is reformulated as a QUBO, it can be solved without using a method specialized to the domain of the problem, and usually produces solutions whose quality rivals that of the specialized methods. 
As discussed in \autoref{subsec:qubo-coefficients-implementation}, we use simulated annealing, but there are many other choices (see, for example, the methods summarized in sections 7 and 8 in \cite{Burkard2013}).

Qubit allocation is an instance of the quadratic assignment problem, meaning our binary variables indicate whether or not to make a particular assignment of logical to physical qubits.
To be explicit,
\begin{equation*}
 x_{ij} = 
    \begin{cases} 1 &\mbox{assign logical qubit }i\hbox{ to hardware qubit }j \\
 0 & \mbox{otherwise }
\end{cases}
\end{equation*}
The problem we now must solve is to find an assignment of logical to physical qubits that minimizes the cost function.
To do so we must choose suitable coefficients (\autoref{subsec:qubo_coefficients}) and enforce any required constraints (\autoref{subsec:qubo-constraints}).


\subsection{QUBO coefficients}
\label{subsec:qubo_coefficients}

In \autoref{eq:qdef}, the coefficients in $Q$ represent relationships between the binary variables.
For qubit allocation, these coefficients should depend on the properties of the circuit in question, and the hardware graph on which we are performing the computation.
The particular form is flexible, and it is an important and defining decision to choose the information and metrics upon which they are created.

QUBO cost functions are often re-expressed as a summation:
\begin{equation}
 \min_x \sum_{ijkl, ij \neq kl} Q_{ijkl} x_{ij} x_{kl} + \sum_{ij} b_{ij} x_{ij}.
 \label{eq:qubo_as_sum}
\end{equation}
The first set of terms are quadratic terms, and their value will relate to the quality of assigning logical qubit $i$ to hardware qubit $j$, and logical qubit $k$ to hardware qubit $l$ in the same allocation.
The diagonal terms of this sum have been removed and re-written as linear terms (since $x_{ij}^2 = x_{ij}$ for binary variables).
The linear terms pertain to the assignments `in isolation', meaning just the consequences of assigning logical qubit $i$ to hardware qubit $j$.

As the QUBO expression represents a cost, the coefficients must be chosen such that  minimization of \autoref{eq:qubo_as_sum} is meaningful.
Given a circuit and hardware graph, we focus on the number and type of one and two qubit gates, the error-rates of the hardware, and the connectivity of the hardware.\footnote{The formulation enables one to easily incorporate other features such as pulse schedules or gate timings, though these are not investigated here.}
Roughly, the coefficients are a product:
\begin{equation}
 f_{\hbox{error}}(\varepsilon) \times f_{\hbox{gate}}(g)  \times f_{\hbox{dist}}(d)
 \label{eq:rough_coef}
\end{equation}
where  $\varepsilon$ is some function of error rates (in our case, we use the success probability),  $g$  a number of gates, and $d$ a distance.
The functions $f_{\hbox{gate}}, f_{\hbox{error}}$, and $f_{\hbox{dist}}$ are then to be determined.
A variety of coefficient forms were tested to determine one that yielded the best results according to a set of metrics: the number of added SWAPs, and the success probability of the final circuit (including any added SWAPs).
These are discussed in detail in \autoref{subsec:qubo-coefficients-implementation}.


\subsection{Handling constraints}
\label{subsec:qubo-constraints}

\begin{figure}[ht]
\centering
\includegraphics[width=.4\textwidth]{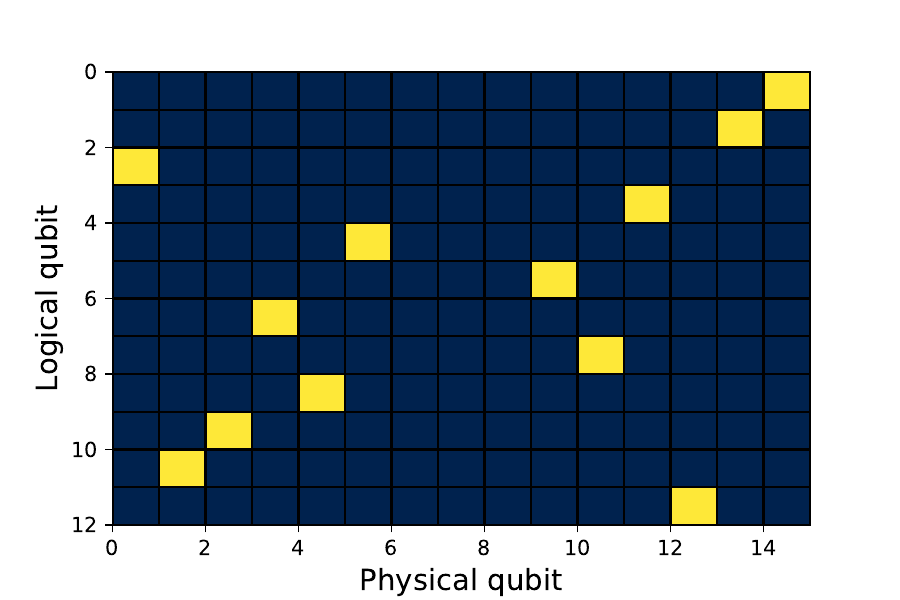}
\caption{An example of a valid allocation for a 12-qubit circuit to a 15-qubit hardware graph. 
         Yellow squares represent an allocated qubit ($x_{ij} = 1$). The constraints of the problem prevent having more than one allocation within a row and column (no under or over assignments of qubits).}
\label{fig:allocation_example}
\end{figure}

The form in \autoref{eq:qubo_as_sum} doesn't explicitly account for constraints on the variables.
For the qubit allocation problem - and quadratic assignment problem in general - allocations that assign the same logical qubit to multiple hardware qubits (as well as assignments of multiple logical qubits to a given hardware qubit) are not valid allocations. 
Mathematically, this is expressed as:
\begin{align}
    \sum_{i=1}^{n_{c}} x_{ij} = 1 \qquad j = 1,...,n_{p} \\
    \sum_{j=1}^{n_{p}} x_{ij} = 1 \qquad i = 1,...,n_{c}
\end{align}
where $n_{c}$ is the number of logical qubits and $n_{p}$ is the number of available hardware (physical) qubits.
In other words, the qubit mapping must be bijective (i.e. a one-to-one mapping).
Visually this can be seen in \autoref{fig:allocation_example}, where no row or column sees more than one allocation.

Constraints  are incorporated by adding penalty terms to the QUBO:
\begin{align}
    \phi \left( \sum_{i=1}^{n_{c}} x_{ij} - 1 \right)^2, \qquad j = 1,...,n_{p} \\
    \theta \left( \sum_{j=1}^{n_{p}} x_{ij} - 1 \right)^2, \qquad i = 1,...,n_{c}
\end{align}
These terms are designed such that for a constraint-violating solution they produce a positive value (thus increasing the cost), but evaluate to 0 if the constraints are satisfied. 
Here $\phi$ and $\theta$ are \emph{penalty coefficients}.
These coefficients control the relative tendency of the optimizer to want to minimize the cost of \autoref{eq:qdef} versus satisfying the constraints.
The specific choice of these coefficients will be discussed in \autoref{subsec:penalty}.
Adding the constraints to \autoref{eq:qubo_as_sum} yields:
\begin{equation}
    \sum^{n_{c}}_{\substack{i=1 \\ i\neq k}}\sum^{n_{p}}_{\substack{j=1 \\ j\neq l}}\sum^{n_{c}}_{k=1}\sum^{n_{p}}_{l=1}Q_{ijkl}x_{ij}x_{kl}
    + \sum^{n_{c}}_{i=1}\sum^{n_{p}}_{j=1}(b_{ij} - (\phi + \theta))x_{ij}
\end{equation}
as the full QUBO for qubit allocation.


\section{Implementation details}
\label{sec:implementation}

We implemented our methods in Python, and make the code available open source on our Github \cite{github}.
An end-to-end example can be found in a Jupyter notebook under `\texttt{examples}' in the Github.

Simulated annealing  was used to find the solutions to our QUBO model.
Specifically, we leverage D-Wave's \texttt{neal} Python package \cite{Dwave}, as it has built-in support for QUBO applications.
Simulated annealing allows us to easily produce distributions of allocations by performing multiple anneals.
This will produce allocations of varying quality, and enables us to see which properties of an initial allocation actually end up mattering for the compiled circuit.
In this context, quality refers to how good the final properties of a fully routed circuit are, as a function of the initial allocation.
The success metrics considered are the number of SWAPs after routing, and the success probability of the final routed circuit.


\subsection{Choosing QUBO coefficients}
\label{subsec:qubo-coefficients-implementation}

As per \autoref{eq:rough_coef}, we would like to construct coefficients that incorporate information about the number of gates, distance between qubits on the hardware graph, and error rates of the gates.
To that end, for the linear terms we would like a coefficient with the form:
\begin{equation}
    b_{ij} = f_{\hbox{error}}(p_{j}) \cdot f_{\hbox{gate}}(g_{i})
    \label{eq:coef-linear}
\end{equation}
where $g_i$ is the number of single-qubit gates acting on qubit $i$, and $p_j$ is the success probability for a single qubit gate on hardware qubit $j$. 
For the quadratic terms, we suppose:
\begin{equation}
    Q_{ijk\ell} = f_{\hbox{error}} (p_{j\ell}) \times  f_{\hbox{gate}}(g_{ik}) \times f_{\hbox{dist}} (d_{j\ell}).
    \label{eq:coef-quadratic}
\end{equation}
Here $g_{ik}$ is the number of two-qubit gates acting on logical qubits $i$ and $k$.
The value of $p_{j\ell}$ is the success probability of executing a two-qubit gate between hardware qubits $j$ and $\ell$ (accounting for any SWAPs that must be added).
Finally $d_{j\ell}$ is the minimum distance between hardware qubits $j$ and $\ell$ on the hardware graph. 
We note that the functions $f$ need not be the same between \autoref{eq:coef-linear} and \autoref{eq:coef-quadratic}.

Computing $p_j$ and $g_i$ for the single qubits is straightforward.
For the two-qubit gates, the coefficients must take into account that for two-qubit gates between hardware qubits that are not connected, SWAPs must be inserted to satisfy these constraints.
This affects how the value of $p_{j\ell}$ is calculated.
In an effort to investigate the quality of just the initial allocation and not the subsequent circuit synthesis and routing process, we `naively' calculate SWAP counts as to not rely on any specific compiler. 
For each two-qubit gate that requires at least one SWAP, we calculate the smallest number of SWAPs that must be added, and assume the qubits are swapped back immediately afterwards\footnote{In \autoref{subsec:nonqueko}, where we present more formal results, the number of SWAPs presented is computed using a proper routing procedure.}.
This number of SWAPs is then used to compute the success probability of a given two-qubit operation.

\begin{figure}[ht]
\centering
\includegraphics[width=.5\textwidth]{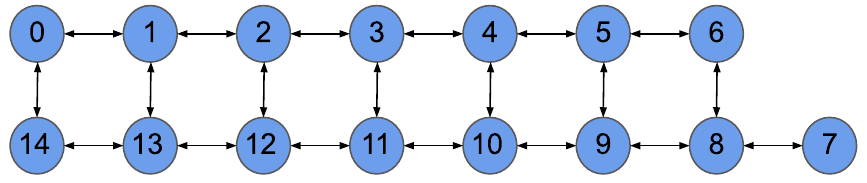}
\caption{Architecture graph for IBM Melbourne, a 15-qubit quantum computer. Edges between nodes indicate qubit connectivity, where each edge is bidirected (can support a CX gate in either direction).}
\label{fig:melbourne}
\end{figure}

Various functions of the quantities of interest were compared using a set of 157 benchmark circuits (taken from \cite{Zulehner2018-2}'s \href{https://github.com/iic-jku/ibm_qx_mapping.git}{Github}) which range from 3-16 logical qubits.
A small amount of filtering had to be performed on this set for this portion of the work.
First, we chose to use the IBM Melbourne hardware graph (\autoref{fig:melbourne}), and so only circuits with up to 15 logical qubits were used.
To gauge the quality of the coefficient forms, we analyzed the percentage difference in naive SWAP count, so for purposes of comparison we do not consider three circuits in which no SWAPs needed to be added.
We also considered a percentage difference of success probability, however for some very large circuits (27 of them), the success probabilities were effectively 0, and these data points were also not used in the coefficient form comparison.

\begin{figure}[ht]
\centering
\includegraphics[width=.5\textwidth]{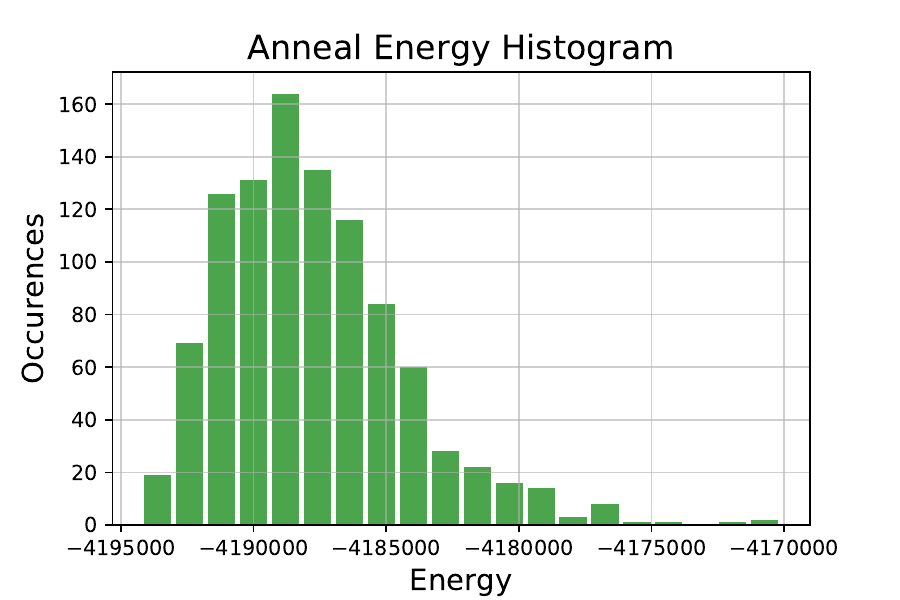}
\caption{ Histogram of 1000 anneal energies for benchmark circuit \texttt{hwb6\_56}, using the coefficient form in \autoref{eq:final-coef}. As sample number increases, these histograms tend to approach a log-normal distribution. The general trend in energies is that for smaller circuits the plot will be very skewed towards lower energies, as the `best' initial allocation (in terms of fewer added SWAPs) will be found for the majority of anneals, while for larger circuits the sample space is too large to find convergence on a particular allocation, and the energies become more normally distributed.}
\label{fig:anneal_hist}
\end{figure}

For each coefficient form and circuit, we performed 1000 anneals and analyzed the resulting allocations.
As a first example, \autoref{fig:anneal_hist} shows the distribution of costs (typically called `energies') for 1000 anneals of a 7-qubit circuit (\texttt{hwb6\_56}) with 3771 single-qubit gates and 2952 two-qubit gates, embedded on a 15-qubit hardware graph (IBM Melbourne, \autoref{fig:melbourne}).
The obtained distribution of energies allows us to verify the behaviour of the QUBO cost function for different coefficient forms, and ensure that our quality metrics are well-correlated with the annealing outcomes.

\begin{figure}[ht]
\centering
\includegraphics[width=.5\textwidth]{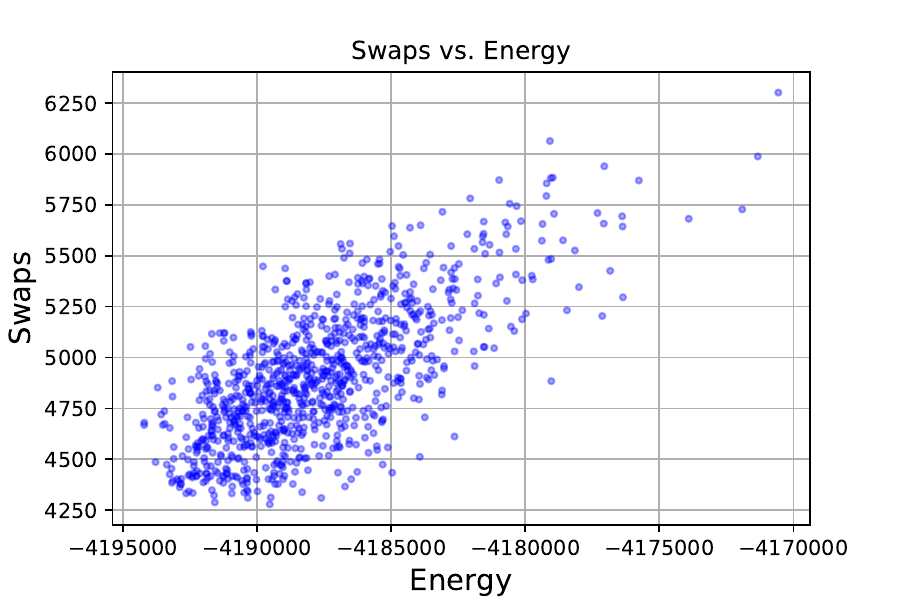}
\caption{Comparison of SWAP count and allocation energy for a 1000 sample anneal run for benchmark circuit \texttt{hwb6\_56}, using the coefficient form in \autoref{eq:final-coef}. SWAP count here is computed using the naive method used to generate the coefficients, without further routing. While there is significant variation, there is a general tendency for lower-energy allocations to also have lower SWAP counts, indicating that the choice of cost function is able to produce quality solutions.}
\label{fig:metric_comp}
\end{figure}

One such comparison is in \autoref{fig:metric_comp}, where we plot the naive SWAP counts for all allocations of a benchmark circuit (\texttt{hwb6\_56}) against their energy for a particular run of 1000 anneals.
Clearly the lower energy allocations tend to also have lower SWAP counts. 
This trend generally held true for all of our benchmark circuits, for both of our quality metrics.

\begin{figure}[ht!]
    \centering
    \subfloat{\includegraphics[width=0.5\textwidth]{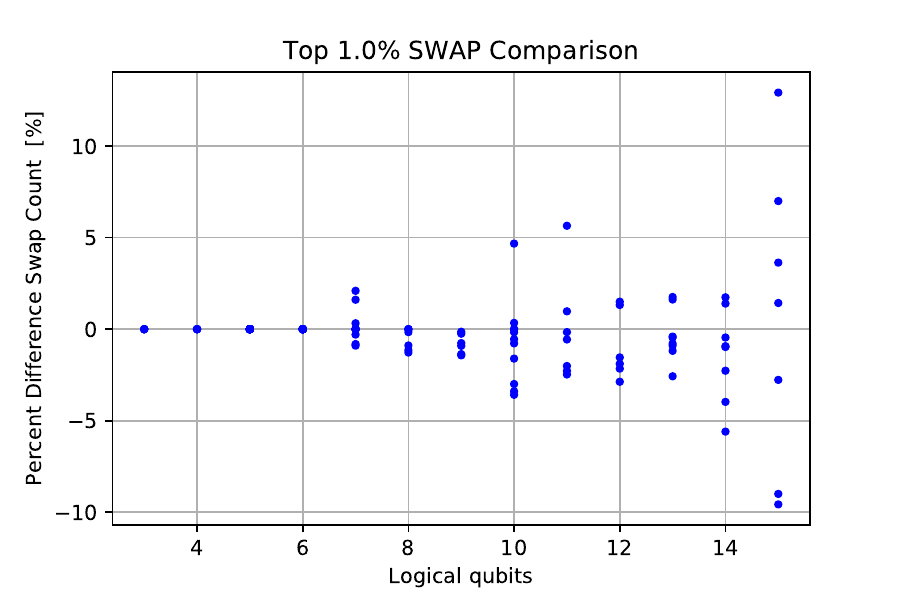}}
    \newline
    \subfloat{\includegraphics[width=0.5\textwidth]{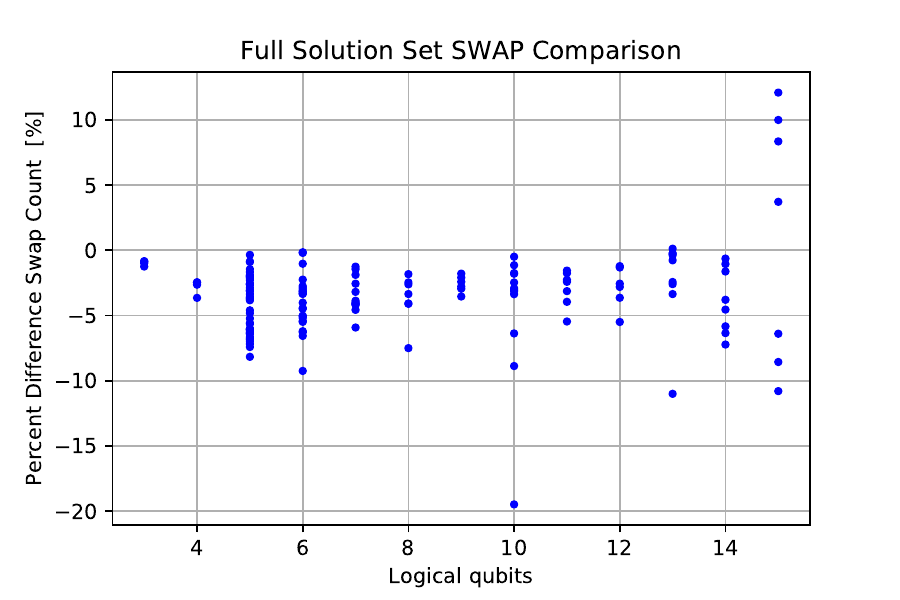}}
    \caption[]{A comparison of the two different QUBO coefficient forms in \autoref{eq:metric_test_notsquared} and \autoref{eq:metric_test_squared}. The quadratic coefficients differ, with one form incorporating distance as-is, while the other uses the distance squared. The plot shows the percentage difference between the two forms' average SWAP counts (over 1000 anneals) over the set of benchmark circuits. A negative value indicates that the squared-distance coefficient form has lower SWAP counts. One sees that for the lowest energy solutions (top), the performance is comparable, with the squared-distance having a slight advantage, but for the distribution as a whole the squared-distance yields consistently lower SWAP counts.}
    \label{fig:coef_comp_1}
\end{figure}

\autoref{fig:coef_comp_1} presents a percentage difference comparison between two candidate forms over the benchmark set:
\begin{equation}
 Q_{ijk\ell} = -\ln(p_{j\ell})  \cdot g_{ik}  \cdot d_{j\ell},\quad b_{ij} = -\ln(p_{j}) \cdot g_{i},
 \label{eq:metric_test_notsquared}
\end{equation}
versus
\begin{equation}
 Q_{ijk\ell} = -\ln(p_{j\ell})  \cdot g_{ik}  \cdot d_{j\ell}^2,\quad b_{ij} = -\ln(p_{j}) \cdot g_{i}
 \label{eq:metric_test_squared}.
\end{equation}
The quantities here are as defined below \autoref{eq:coef-quadratic}.
The top plot shows the percentage difference of average naive SWAPs from the top 1\% of allocations for each form (i.e. the lowest energy solutions), and the bottom plot the percentage difference of naive SWAPs for the full set of allocations.
For the top 1\% we see that both coefficient forms find the lowest SWAP allocation for smaller circuits, but start to diverge as the number of logical qubits increases past 6, where the form with $d_{j\ell}^2$ clearly finds better allocations over the one with just $d_{j\ell}$.
In the plot that considers the full set of allocations, there is no convergence for smaller circuits and it is apparent that the form in \autoref{eq:metric_test_squared} is superior for the vast majority of benchmark circuits.

\begin{figure*}[ht!]
\centering
\includegraphics[width=\textwidth]{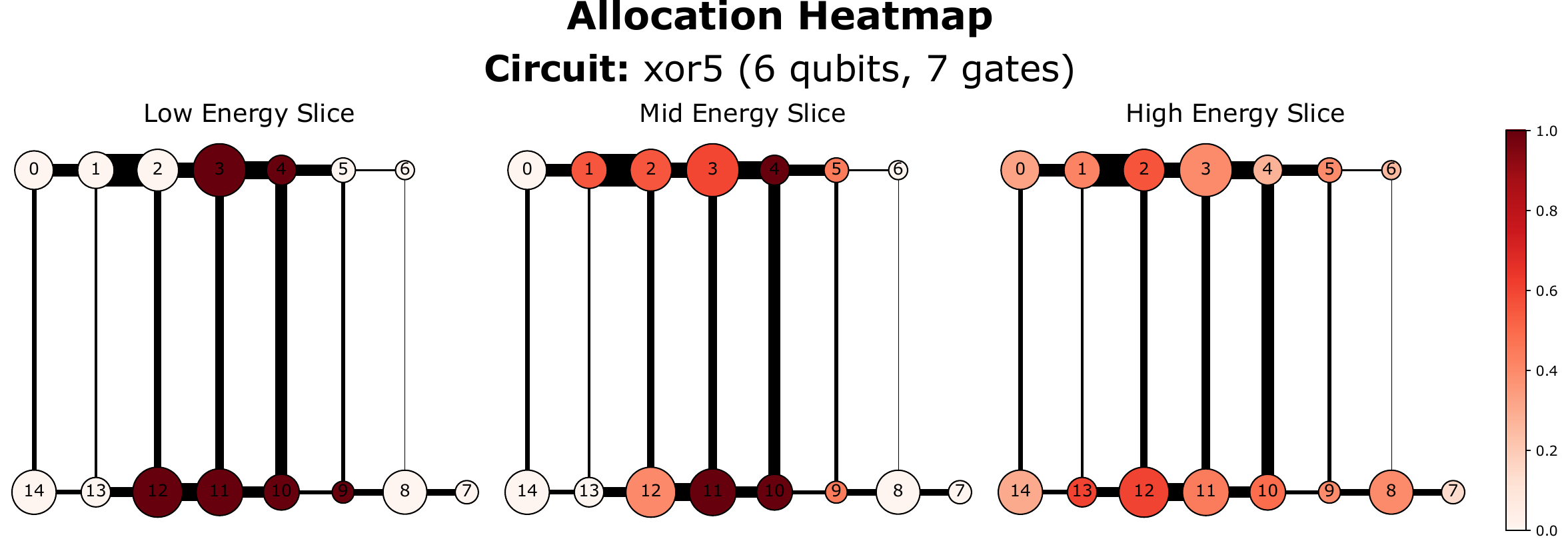}
\caption{Heatmap showing a range of allocations to the IBM Melbourne hardware graph obtained from a 1000-sample run of simulated annealing. Each panel shows 5$\%$ of the solutions for a given energy range. Darker colour indicates a higher concentration of allocations involving that specific qubit. The node and edge sizes are proportional to one and two-qubit error rates, where bigger is better (i.e. smaller error-rates). Lower-energy (i.e. higher quality) solutions shown on the left see the allocations clustering around the most well-connected qubis with the lowest error rates, whereas higher-energy solutions are more variable.}
\label{fig:heatmaps}
\end{figure*}

To look deeper into the structure of the allocation process we used heatmaps, where different sets of allocations at different energies are plotted on the hardware graph to show the distribution of the qubits being assigned.
In \autoref{fig:heatmaps} are heatmaps for IBM Melbourne representing the lowest 5\% of the energies, the middle 5\%, and the highest 5\%, taken from a 1000-sample annealing run.
The qubits are coloured based on the fraction of allocations within that energy range that include those qubits.
The node and edge sizes are scaled based on the single and two-qubit error rates respectively, where a larger size means a better (lower) single-qubit error-rate, and thicker edges indicate better (lower) two-qubit error rates. 
We notice that in these examples, for low energies the allocations tend to converge on the most well-connected qubits with the lowest error-rates, dispersing as the energies increase.
This is visual affirmation that lower energy allocations are better allocations, as they converge on the `best' available physical qubits.

We continued testing various coefficient forms using the percentage difference comparison method (the full complement of which can be found on our Github \cite{github}) ultimately concluding that the form:
\begin{equation}
    Q_{ijk\ell} = -\ln(p_{j\ell})  \cdot g_{ik}  \cdot d_{j\ell}^{3},\quad b_{ij} = -\ln(p_{j}) \cdot g_{i}
    \label{eq:final-coef}
\end{equation}
generally performs well based on our quality metrics.
It is worth noting that squaring the graph distances (as in \autoref{eq:metric_test_squared}) actually performed better {on average for smaller circuits (in terms of logical qubit number), but worse for larger circuits.
Perhaps with further investigation, one could find a threshold to decide which exponent to use based on input circuit properties.

While this is the best coefficient form we found, we encourage the interested reader to investigate other forms, or other quality metrics that may better suit their purposes.
In general we found that incorporating a particular metric as a part of the QUBO coefficients will improve the final solutions of the QUBO with respect to that metric.
During the experimentation process we determined some good rules of thumb for this inclusion.
For example, multiplying the desired metrics performed better than adding them, and allocations obtained by taking the natural log of the success probabilities produced routed circuits with higher success probabilities.
We also tried re-scaling the number of one- and two-qubit gates, removing the linear term ($b_{ij}$) entirely, and using subsets of the three metrics shown in \autoref{eq:coef-quadratic}, however none of these produced allocations with significantly better quality than \autoref{eq:final-coef}.

As a final note, one question that arises is whether correlation with energy is present after the circuit has been compiled (and thus fully routed) to insert any necessary SWAPs.
A test of this is shown in \autoref{fig:qiskit_distribution}, where we take a sample of 1000 allocations from an anneal run for a 14-qubit circuit (\texttt{cm42a\_207}) with 1005 single-qubit gates and 771 two-qubit gates, using IBM Melbourne as a hardware graph.
Each allocation is given to Qiskit's compiler (v0.20.0) using level 0 optimization and the `basic' routing method, and the SWAP counts of the compiled circuits are plotted.
As can be seen visually, there is no strong correlation between QUBO allocation energies and compiled circuit SWAP counts. 
This means that we cannot predict which QUBO initial allocation will produce the best compiled circuit.

\begin{figure}[ht]
\centering
\includegraphics[width=.5\textwidth]{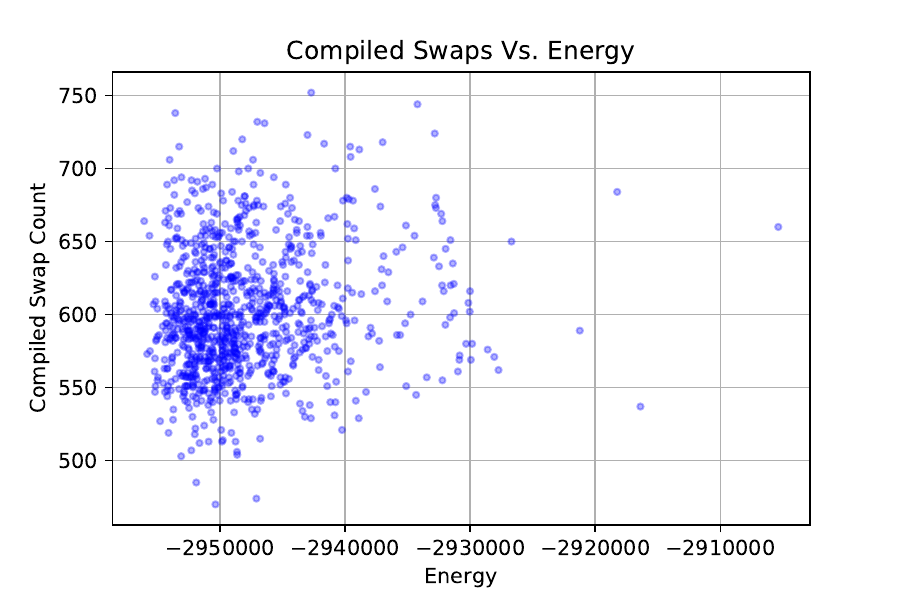}
\caption{Distribution of Qiskit SWAP counts for each allocation in a 1000 sample anneal run for benchmark circuit \texttt{cm42a\_207}. Compiled using Qiskit v0.20.0 using a preset passmanager with optimization level 0 and the `basic' routing method. We can infer that it is not possible to predict the compiled circuit SWAP counts based on the energy of a given allocation.}
\label{fig:qiskit_distribution}
\end{figure}

It would be of particular interest to further investigate other QUBO coefficient forms that would have such predictive power.
We leave this as an interesting problem for future work.
We can still be confident that the lower energy QUBO allocations are better initial allocations from the fact that they are well correlated with our own quality metrics, meaning the QUBO method is self-consistent.
Furthermore, comparison must also be made against other initial placement techniques to investigate if there are improvements from the QUBO allocations, as even if the compiled SWAPs are not so correlated, we may still see improvements on average.
Highlights of such comparisons are made in \autoref{subsec:nonqueko}, with the full set of results in Appendix~\ref{app:results}.


\subsection{Penalty coefficients}
\label{subsec:penalty}

In addition to choosing a form for $Q_{ijk\ell}$ and $b_{ij}$, we must also choose values of $\phi$ and $\theta$.
This aspect is often not discussed in the literature for other QUBO applications, which simply specifies that a typical choice is 75-150\% of the maximum value of the coefficient matrix without constraints added (see end of section 4.1 in \cite{Glover2018}).

We found that the above rule-of-thumb worked reasonably well, but required a small amount of tweaking.
The final process for deciding penalty values consisted of first setting both $\phi$ and $\theta$ equal to the maximum coefficient matrix value.
We would then check if any of the returned allocations from simulated annealing did not satisfy the constraints.
If at least one was invalid, we would re-run that circuit until all the allocations satisfied the constraints, multiplying the penalties by 2, then 3, etc. on each successive run.
In the course of our benchmarking, no circuit ever had to have their penalties increased beyond 3 times the maximum coefficient matrix value, with the vast majority of circuits succeeding without needing to be re-run.

As a final point of interest, we had initially included an additional constraint,
\begin{equation}\label{gampen}
    \gamma \left(\sum^n_{i=1}\sum^n_{j=1}x_{ij} - n_c \right)^2,
\end{equation}
which ensured that the correct number of qubits were assigned.
The concern was that for circuits where the number of logical qubits was less than the amount of available hardware qubits, it would tend to assign more qubits than required.
As it turned out, adding this additional constraint was unnecessary --- the other constraints appear to implicitly handle this --- and actually made it more challenging to set the penalty coefficients.

We note that given the results of some initial tests while hand-tuning penalty coefficients, it does seem like there is an optimal range for each circuit (paired with a particular coefficient form).
An interesting topic of future investigation could be to automate the selection process, or employ something akin to a Newton's method algorithm to converge on these circuit-optimal ranges.


\section{Benchmarks}
\label{sec:benchmarks}

In this section we analyze the effectiveness of the QUBO formulation and compare its performance to other contemporary allocation methods available in \tket (py\tket v0.5.7) \cite{Sivarajah2020} and Qiskit (v0.20.0) \cite{Qiskit} (\autoref{subsec:nonqueko}).
In \autoref{subsec:queko} we also analyze performance with respect to the set of recently-proposed QUEKO benchmarks \cite{Tan2020}.
All benchmarks were run on a machine with an Intel i5-4590 4-core processor at 3.30 GHz with 16 GB of RAM.

\subsection{Comparison against contemporary initial placement methods}
\label{subsec:nonqueko}

Using the benchmarks available at \cite{Zulehner2018-2}'s \href{https://github.com/iic-jku/ibm_qx_mapping.git}{Github}, the performance of our method was compared to initial allocation methods available in \tket and Qiskit.
We also compare with the initial allocation method used by the SABRE algorithm \cite{Li2018}, which was recently added to Qiskit.

For these benchmarks, we use IBM Melbourne (see \autoref{fig:melbourne}) as our hardware-graph, and for each circuit we take the QUBO allocation with the lowest naive-SWAP count.
As we care primarily about the performance of the initial allocation, we specifically turn off all further circuit synthesis optimization so that the Qiskit and \tket compilers are essentially just routing the mapped circuits.
For Qiskit, this corresponds to using the level 0 optimization preset passmanager with the `basic' router.
For \tket we set all routing parameters available in the \texttt{router} function (\texttt{swap\_lookahead}, \texttt{bridge\_lookahead}, \texttt{bridge\_interactions}, \texttt{bridge\_exponent}) to 0 and didn't apply any further optimization calls.

Before calculating any of the compiled circuit's properties, we decompose all SWAP gates to CX gates (for \tket we also decompose all BRIDGE gates).
To assess the quality of the initial allocation, we compute the total number of CX gates and the circuit depth for the routed circuit, as these are (generally) the metrics that most other qubit allocation methods report.

As for our choice of benchmark circuits, since we are allocating to a 15-qubit hardware, we must ignore 6 of the 157 benchmark circuits due to them using 16 logical qubits, leaving us with 151 benchmarks.
For \tket we use all 151 remaining benchmarks, but for Qiskit we restrict ourselves to using only circuits with 10,000 or fewer total gates, due to poor time-scaling for very large circuits.
With this restriction in place, we are left with 131 benchmark circuits for comparison with Qiskit.
The detailed results are included in Appendix~\ref{app:results} for both compilers.
In this section we discuss the highlights.

\begin{table}
\begin{center}
\begin{tabular}{ l || c | c }

 Allocation Method & CX count [\%] & Depth [\%]  \\
 \hline
 \texttt{LinePlacement} & 56.9 & 58.9 \\  
 \texttt{GraphPlacement} & 55.0 & 59.0 \\  
 \texttt{Trivial} & 90.1 & 85.1 \\
 \texttt{Dense} & 68.6 & 70.2\\
 \texttt{Noise} & 57.9 & 77.0 \\
 \texttt{SABRE} & 48.8 & 53.7\\

\end{tabular}
\caption{Table showing the percentage of total benchmark circuits for which QUBO demonstrated improvement over the specified allocation methods, for both performance metrics (total CX count and circuit depth). The first two methods are from \tket and the subsequent four from Qiskit.}
\label{tab:overall-results}
\end{center}
\end{table}

For \tket we compare to the \texttt{LinePlacement} and \texttt{GraphPlacement} initial allocation methods.
For Qiskit we compare to the \texttt{Trivial}, \texttt{Dense}, \texttt{Noise} and \texttt{SABRE} allocation methods.
In \autoref{tab:overall-results} we present the overall performance of the QUBO method in terms of the percentage of benchmark circuits in which it yielded improvement over the other initial placement methods.
Notably, QUBO finds a better allocation for $>$50\% of the benchmark circuits compared to almost every other allocation method, in terms of CX count and circuit depth. The single exception is that the SABRE placement technique more often yields better CX counts.
While this big picture result is valuable, we must analyze the performance on a more fine-grained level.

\begin{figure}[ht]
\centering\
\includegraphics[width=.5\textwidth]{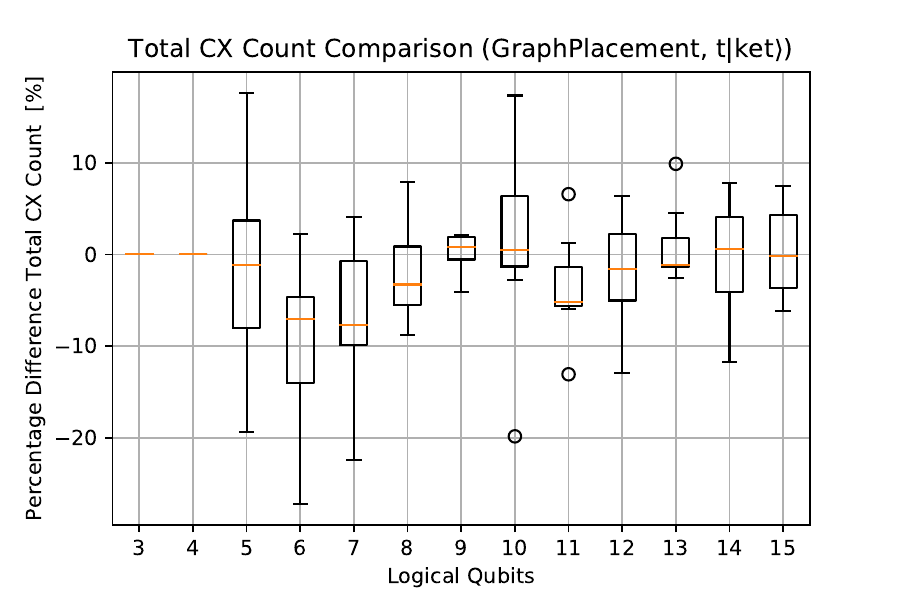}
\caption{Box plot of the percent difference comparison between the \tket \texttt{GraphPlacement} initial allocation method and the QUBO lowest SWAP allocation for total CX count, over all applicable benchmark circuits. The difference is taken such that a negative value indicates that the QUBO-obtained allocations required fewer added CX gates.}
\label{fig:graph_cx_comp}
\end{figure}

\begin{figure}[ht]
\centering\
\includegraphics[width=.5\textwidth]{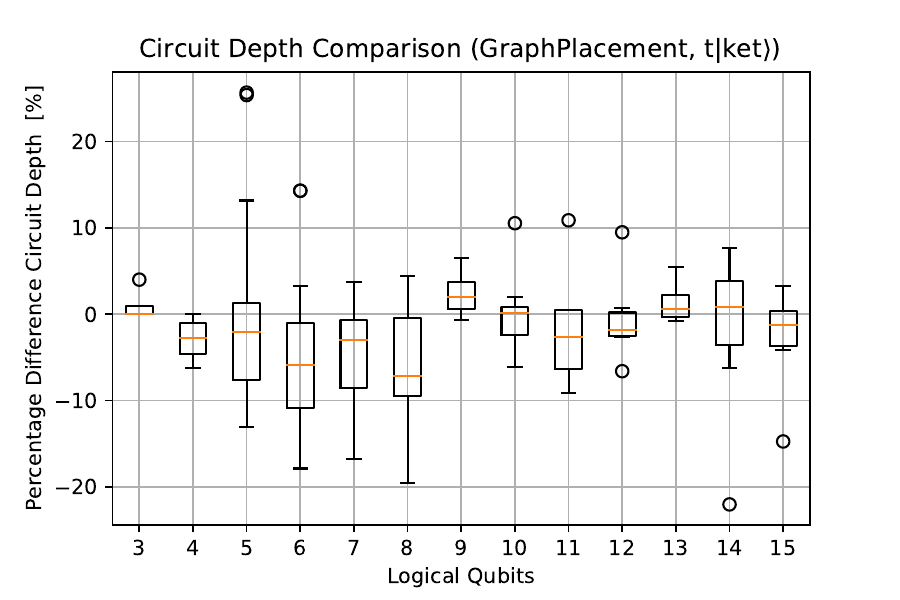}
\caption{Box plot of the percent difference in circuit depth between the \tket \texttt{GraphPlacement} initial allocation method and the QUBO lowest naive-SWAP allocation over all applicable benchmark circuits. A negative value indicates that the QUBO-obtained allocations had a lower circuit depth.}
\label{fig:graph_depth_comp}
\end{figure}

To compare the initial allocation's compiled circuit properties (total CX count and circuit depth) we again employ percent-difference comparisons, plotting the results as box-and-whisker plots as a function of logical qubit number.
\autoref{fig:graph_cx_comp} and \autoref{fig:graph_depth_comp} show the comparison between the QUBO method and the \tket \texttt{GraphPlacement} method for total CX count and circuit depth, respectively.
The overall performance of both methods is fairly circuit dependent, but the distributions seem to be skewed in QUBO's favour, especially for smaller circuits.
In the smallest circuits (four or fewer logical qubits) both methods converge on the same CX counts, but still differ slightly in circuit depth, with a very slight edge to QUBO.
While we don't show here a percentage difference comparison  between \texttt{LinePlacement} and QUBO (it is available on our Github \cite{github}) we note that it is fairly similar to the \texttt{GraphPlacement} comparison, with the only notable difference being that for two of circuits, the QUBO allocation obtained much lower depth and CX count ($>$55\% difference for depth, $>$40\% difference for CX count).
We also note that the \texttt{GraphPlacement} method was significantly slower at finding allocations compared to both \texttt{LinePlacement} and QUBO, taking several minutes for some circuits in this benchmark set, and in some cases hours for some circuits in the to-be-discussed QUEKO B$_{SS}$ benchmark set (see \autoref{subsec:queko} for more detail.)

For comparison with Qiskit, the methods of interest are the \texttt{Dense} and \texttt{SABRE} allocation methods.
Qiskit also contains two additional methods, \texttt{Trivial} and \texttt{Noise}, but \texttt{Noise} performs similarly to \texttt{Dense}, and unsurprisingly the \texttt{Trivial} method performs poorly compared to all the other methods.

In \autoref{fig:dense_stacked_cx} and \autoref{fig:dense_stacked_depth}, we can see the comparison between QUBO and \texttt{Dense}. 
Looking at the distributions, most are skewed in favour of the QUBO allocation, partly due to some outlier circuits where QUBO performed significantly better than \texttt{Dense}.
We see again that the QUBO allocations are more favourable for the smaller circuits in the benchmark set than the larger ones, but this is more likely to be a function of the gate compositions of the circuits than a function of logical qubit number, given that we see such large variation for particular logical qubit numbers.

\begin{figure}[ht]
\centering\
\includegraphics[width=0.5\textwidth]{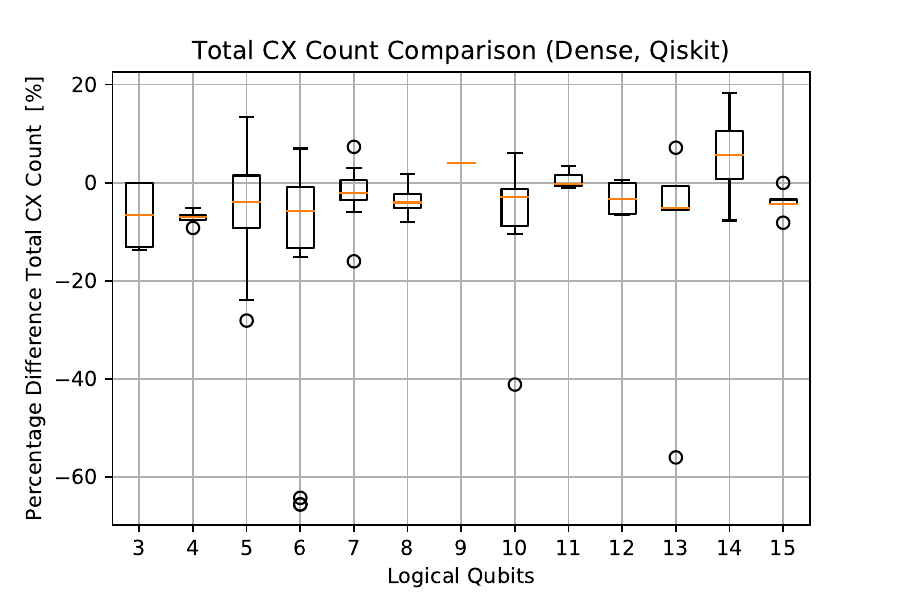}
\caption{Box plots of the percent difference comparison in total CX count between Qiskit's \texttt{Dense} initial allocation method and the QUBO lowest naive-SWAP allocation over all applicable benchmark circuits. A negative value indicates that QUBO allocations had fewer CX gates added.}
\label{fig:dense_stacked_cx}
\end{figure}

\begin{figure}[ht]
\centering\
\includegraphics[width=0.5\textwidth]{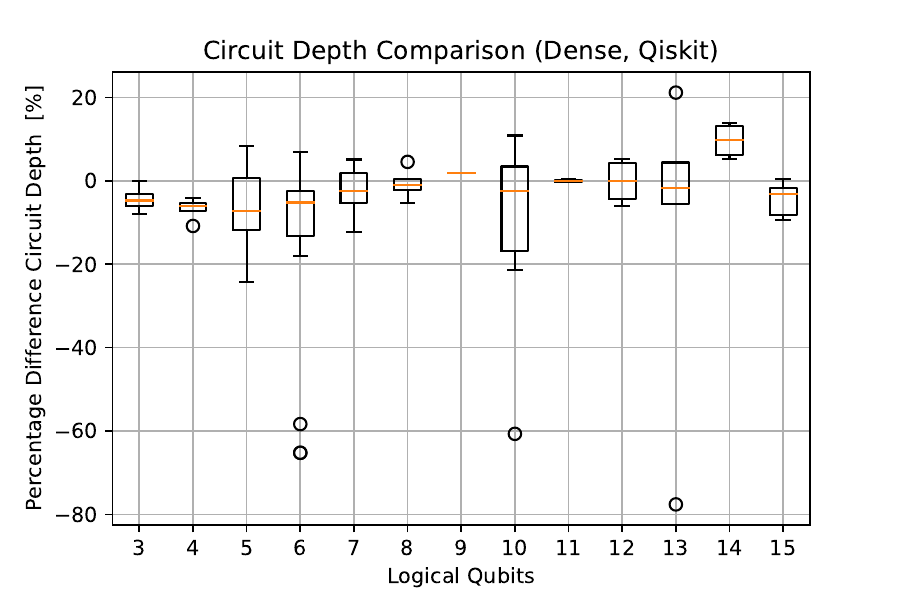}
\caption{Box plots of the \% difference comparison between Qiskit's \texttt{Dense} initial allocation method and the QUBO lowest SWAP allocation for circuit depth, over all applicable benchmark circuits. The difference is taken such that a negative value indicates that QUBO allocations had smaller depths.}
\label{fig:dense_stacked_depth}
\end{figure}

In \autoref{fig:SABRE_stacked_cx} and \autoref{fig:SABRE_stacked_depth} we compare QUBO and \texttt{SABRE}.
Again we see some large outlier circuits for which QUBO does significantly better.
In terms of the distributions of percent differences, both QUBO and \texttt{SABRE} seem to perform equally well for smaller circuits, with a slight edge to QUBO for the medium sized circuits and to \texttt{SABRE} for the larger circuits.
This is more pronounced in the depth plot, where there is a clear dip in the middle.
\texttt{SABRE} seems to also do a slightly better job at finding smaller depths than QUBO.

\begin{figure}[ht]
\centering\
\includegraphics[width=0.5\textwidth]{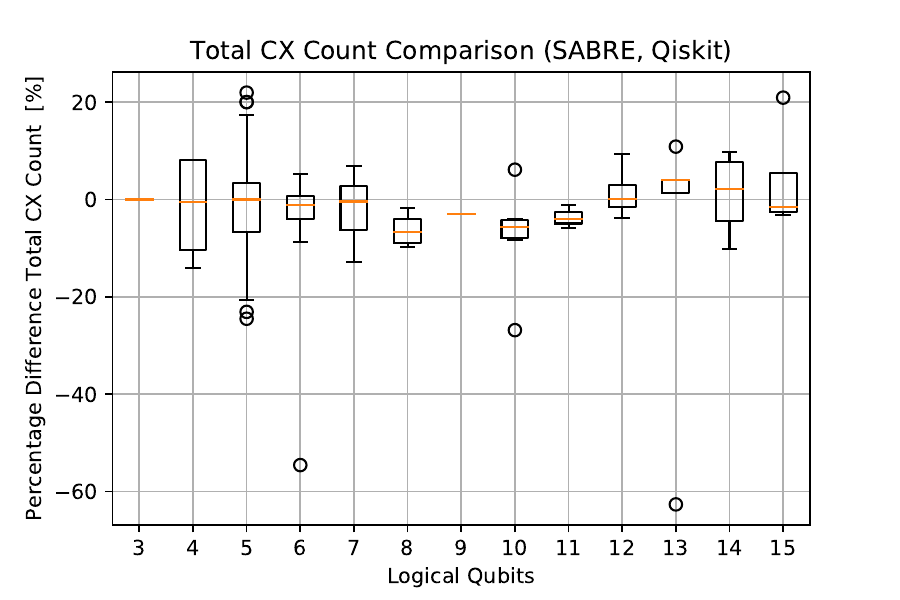}
\caption{Box plots of the \% difference comparison between Qiskit's \texttt{SABRE} initial allocation method and the QUBO lowest SWAP allocation for total CX count, over all applicable benchmark circuits. A negative value indicates that QUBO allocations had fewer CX gates added.}
\label{fig:SABRE_stacked_cx}
\end{figure}

\begin{figure}[ht]
\centering\
\includegraphics[width=0.5\textwidth]{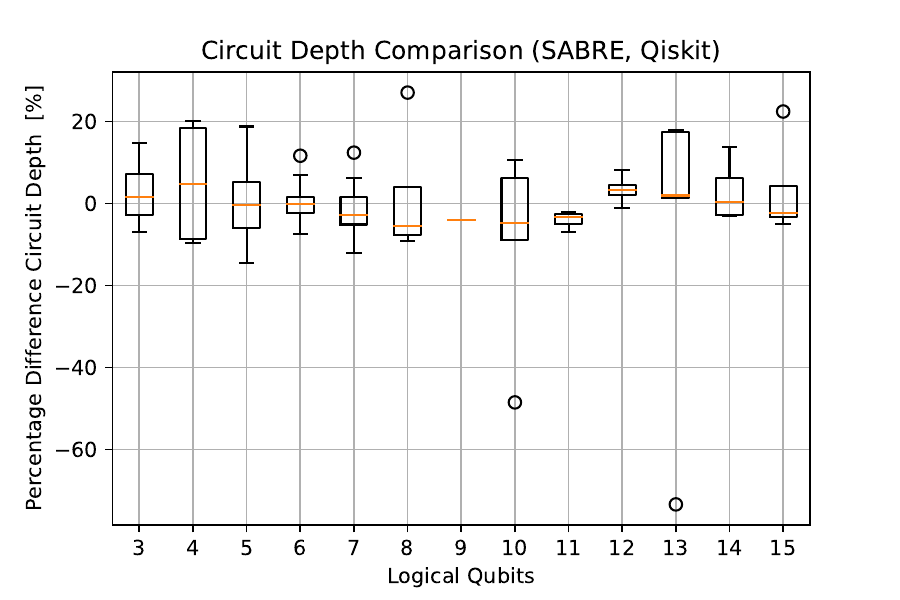}
\caption{Box plots of the \% difference comparison between Qiskit's \texttt{SABRE} initial allocation method and the QUBO lowest SWAP allocation for circuit depth, over all applicable benchmark circuits. A negative value indicates that QUBO allocations had smaller depths.}
\label{fig:SABRE_stacked_depth}
\end{figure}

Our analysis suggests that there is no universally best method among the ones considered.
All the various initial allocation methods seem to do well for particular circuits, as shown by the wide and variable distributions in the plots.
It would be of particular interest to know which features of circuit composition lead to the better performance of one allocation method versus another.
We leave this to future work.

\subsection{QUBO performce for QUEKO Benchmarks}
\label{subsec:queko}

Recently, a set of known optimal-depth benchmark circuits were presented.
In addition to proposing these QUEKO benchmarks, the authors also compared the performance of many different publically available compilers \cite{Tan2020}.
Their benchmarks are broken down into two main sets, one being B$_{NTF}$, or the `near-term feasible' benchmarks, which range from optimal depth 5 to 45, and the other being B$_{SS}$, or the `scaling study' benchmarks, ranging from  optimal depth 100 to 900.
Surprisingly, their results show that even the most competitive compilers available today have trouble getting close to the depth optimal solutions, deviating often to 5x the optimal depth or even greater (though \tket demonstrated remarkable performance, obtaining results very close to optimal).
The benchmarks themselves are produced for a variety of hardware graphs, ranging from 16 qubits (Rigetti's Aspen-4) to 53 qubits (Google's Sycamore), so one can also look at the effect that increasing the space of possible allocations has on the compiler's performance.

We were curious how QUBO initial allocations would do at finding the optimal allocation for these benchmarks, so we fed some QUBO initial allocations through \tket and Qiskit's compiler to check which depths they could achieve.
In this case, we are interested in how well the compiling process does as a whole and so we use the same optimization settings as \cite{Tan2020}.
After running the compilers we decompose any SWAP gates to CX (as well as BRIDGE for \tkets) and then record the realized depth.
Again, we choose the lowest naive-SWAP QUBO allocation for each QUEKO benchmark circuit.
An important point to note is that for these anneals, due to the larger size of some of the hardware}, we only ran 100 samples of annealing in order to minimize runtime.

For most of the hardware used, we did not have access to their calibration data and therefore did not have any success probabilities to include in our QUBO coefficients.
To work around this we removed the success probability term from \autoref{eq:final-coef}, and did a percent difference comparison for this form with and without the probability term using IBM-Melbourne calibration data and the benchmark set used in \autoref{subsec:nonqueko}.
We did this comparison to see if the quality of the allocations differed significantly between the forms.
Unsurprisingly, the coefficent form that did not include success probability did worse at finding circuits with higher success probabilities, but in terms of SWAP counts the percent differences were scattered around 0, meaning both coefficient forms performed comparably.

\begin{figure*}[ht!]
    \centering
    \subfloat[Aspen-4 (16 Qubits)]{\includegraphics[width=0.5\textwidth]{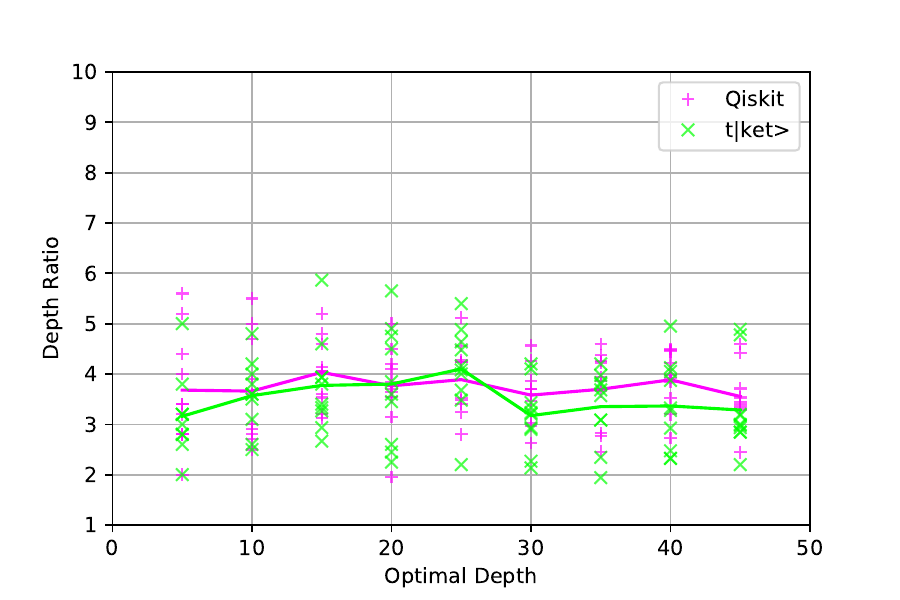}}
    \subfloat[Sycamore (53 Qubits)]{\includegraphics[width=0.5\textwidth]{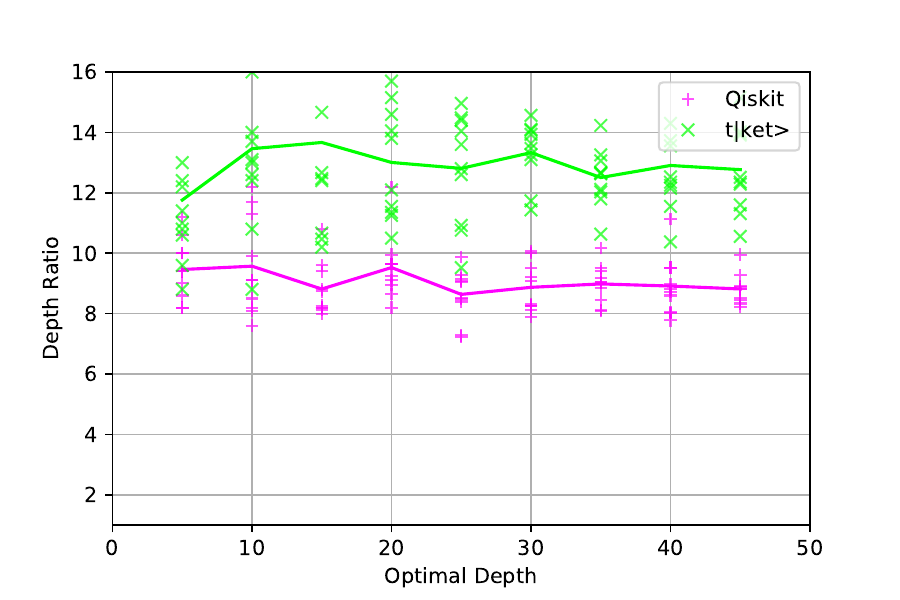}}
    \caption{Performance of a QUBO initial allocation given to t$\mid$ket$\rangle$ and Qiskit for the QUEKO B$_{NTF}$ circuits (B$_{NTF}$ = Benchmarks for near-term feasibility). Each data point is a unique circuit for a specific optimal depth, and lines are 10 circuits averages per depth per hardware-graph (180 circuits total).}
    \label{fig:queko-bntf}
\end{figure*}

\begin{figure*}[ht!]
    \centering
    \subfloat[t$\mid$ket$\rangle$ routing]{\includegraphics[width=0.5\textwidth]{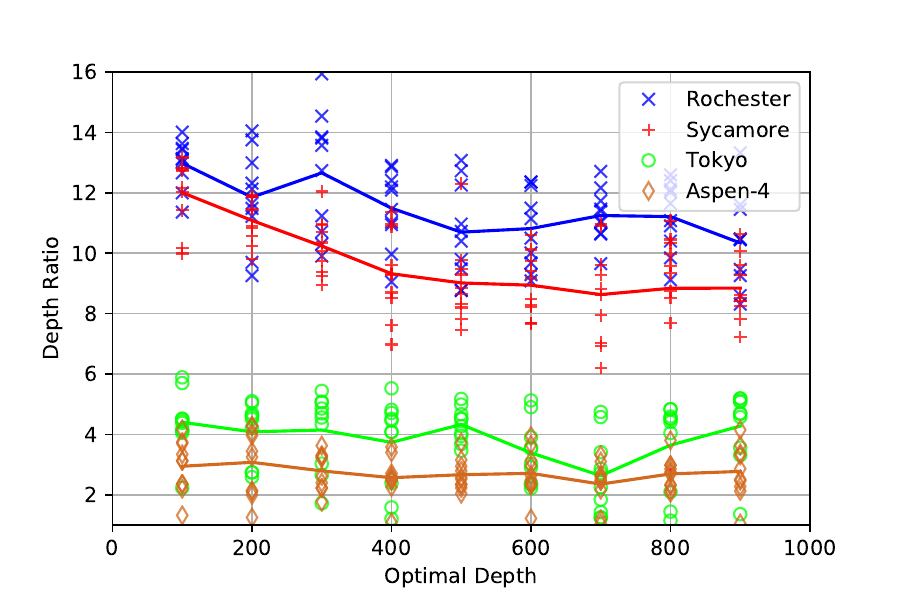}}
    \subfloat[Qiskit routing]{\includegraphics[width=0.5\textwidth]{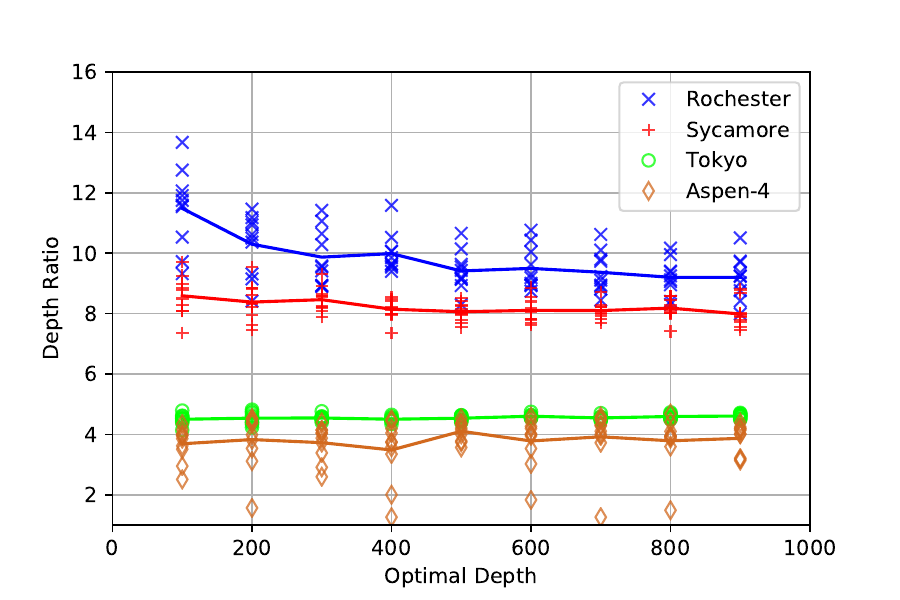}}
    \caption{Performance of a QUBO lowest naive-SWAP initial allocation given to t$\mid$ket$\rangle$ and Qiskit for the QUEKO B$_{ss}$ circuits (B$_{ss}$ = Benchmarks for Scaling Study). Each data point is a unique circuit for a specific optimal depth, and lines are 10 circuits averages per depth per hardware-graph (360 circuits total).}
    \label{fig:queko-bss}
\end{figure*}

In \autoref{fig:queko-bntf} and \autoref{fig:queko-bss} we recreate figures 6 and 7 from \cite{Tan2020}, but using QUBO initial allocations that are given to \tket and Qiskit's compilers.
It's important to detail that we are using a different \tket and Qiskit version than is used in \cite{Tan2020}, so the results are not directly comparable.
This was discovered through some tests courtesy of the authors of \cite{Tan2020}, where it was found that using the newer version of \tket results in significantly different compilation results \cite{Tan_Bochen_2020}.
Therefore our results should be interpreted in the truest sense of how close we get to the optimal depths, and not compared directly to the results present in the QUEKO paper.

Plotted in \autoref{fig:queko-bntf} is a comparison of Qiskit and \tkets's compiled circuits on Aspen-4 and Sycamore for the B$_{NTF}$ benchmark set. 
For Aspen-4, both compilers perform similarly given the same QUBO initial allocation, but diverge for the Sycamore circuits, where Qiskit clearly finds closer-to-optimal depths.
It seems that the size difference of the hardware is the primary contributor to the difficulty of finding the depth-optimal allocations, and not the circuit depth.
This same effect is seen in the B$_{SS}$ benchmark set in \autoref{fig:queko-bss}, where across an order of magnitude in circuit depth, we see very little variation as a function of depth, but a pronounced difference between the hardware being used.
In particular, the smaller hardware sees close to optimal depths while the larger hardware is quite a bit from optimum.
Comparing compilers, neither \tket nor Qiskit had particularly remarkable performance with the QUBO initial allocations, besides a small performance increase for \tket when moving to higher depth circuits initially.

An interesting experiment would be to see whether generating more samples during our anneal runs would combat the effect of the increase of the space of possible allocations as one moves to larger hardware and help in finding the more optimal allocations.
Overall, it seems QUBO struggles with finding optimal depth allocations for larger hardwares and we leave it to future studies to see if some coefficient form or other change could improve on this performance.

\section{Conclusion}
\label{sec:conclusion}

The QUBO formulation has some very useful properties compared to other initial allocation methods.
The coefficient form is very flexible, allowing the consideration of whatever metrics one cares to improve on for the allocation process.
It is also agnostic to gate count --- only limited on the solver end --- therefore able to handle circuits of arbitrary depth (for both single and two-qubit gates).
It is important to note that this method is specifically an initial allocation method, meaning it performs no further circuit optimization past finding good initial allocations.

While we showed that the QUBO method performs at the level of or slightly better than other available initial allocation methods, this just scratches the surface and many optimizations could be investigated.
For example, one could attempt to improve the results by adjusting the simulated annealing hyperparameters.
The results presented here use only the default settings provided in \texttt{neal}, but in general one could tweak the properties of the simulated annealer (e.g. temperature schedule, number of sweeps etc.) in order to produce better allocations.
Furthermore, while we have not done so, it would be interesting to leverage special-purpose annealing hardware (whether classical or quantum) to see if we can obtain performance improvements, in particular for larger circuits with higher numbers of logical qubits.
As mentioned in \autoref{subsec:penalty}, it would also be valuable to investigate a behind-the-scenes optimizer for the penalty coefficient values, as their magnitude did seem to effect allocation quality.
In terms of future directions, it would be beneficial to know if generating more simulated annealing samples improves the quality of allocations when using larger hardware graphs (around the size of Sycamore, 53 qubits).

In general our results suggest that further research is necessary to learn the cause of the large variations seen in the performance of the initial allocation methods.
Given the differences between circuits, it is likely that the circuit composition also plays a large role, most likely dominated by the two-qubit interactions and the necessary routing that must be done to satisfy connectivity constraints.
Studying how the properties of a circuit affect the quality of allocation and routing methods will lead to even more improvements, and in particular for the QUEKO benchmarks will help close the optimality gap.


\section{Acknowledgements}

We thank Bochen Tan and Rodney Van Meter for helpful discussions.
TRIUMF receives federal funding via a contribution agreement with the National Research Council of Canada.
BD acknowledges funding from the TRIUMF student program, BioTalent Canada, and RBC Future Launch. We acknowledge the use of IBM Quantum services for this work to obtain hardware graphs and qubit calibration data. The views expressed are those of the authors, and do not reflect the official policy or position of IBM or the IBM Quantum team.

 \bibliography{main.bib}


\onecolumngrid

\appendix

\section{Benchmark tables}
\label{app:results}

This section contains the full set of data for the plots and benchmarks discussed in \autoref{sec:benchmarks}.

\begin{longtable}{lc|rr||rr|rr|rr}
\caption{Initial placement comparison of QUBO to the methods available in \tket using IBM Melbourne. Reported depths take into account all gates (not just CX). We bold the best performing method in each row (for ties, each tieing method is bolded).}
\label{tket_table} \\

\multicolumn{2}{c}{} & \multicolumn{2}{c}{\textbf{Initial}} & \multicolumn{2}{c}{\textbf{Line}} & \multicolumn{2}{c}{\textbf{Graph}} & \multicolumn{2}{c}{\textbf{QUBO}} \\
 Circuit                    &   n &    Depth &   CX count &   Depth &    CX &   Depth &    CX &   Depth &  CX \\
\toprule
\endfirsthead

\multicolumn{2}{c}{} & \multicolumn{2}{c}{\textbf{Initial}} & \multicolumn{2}{c}{\textbf{Line}} & \multicolumn{2}{c}{\textbf{Graph}} & \multicolumn{2}{c}{\textbf{QUBO}} \\
 Circuit                    &   n &    Depth &   CX count &   Depth &    CX &   Depth &    CX &   Depth &  CX \\
\toprule
\endhead

  ex1\_226                    &   6 &               5 &                  5 &           11 &        11 &\textbf{             7 }&         11 &            8 &\textbf{         8 
}\\ graycode6\_47               &   6 &               5 &                  5 &\textbf{            5 }&\textbf{         5 }&\textbf{             5 }&\textbf{          5 }&\textbf{            5 }&\textbf{         5 
}\\ xor5\_254                   &   6 &               5 &                  5 &           11 &        11 &\textbf{             7 }&         11 &            8 &\textbf{         8 
}\\ ex-1\_166                   &   3 &              12 &                  9 &\textbf{           25 }&\textbf{        21 }&\textbf{            25 }&\textbf{         21 }&           26 &\textbf{        21 
}\\ 4mod5-v0\_20                &   5 &              12 &                 10 &\textbf{           27 }&\textbf{        22 }&\textbf{            27 }&\textbf{         22 }&\textbf{           27 }&\textbf{        22 
}\\ 4mod5-v1\_22                &   5 &              12 &                 11 &\textbf{           27 }&\textbf{        23 }&\textbf{            27 }&\textbf{         23 }&\textbf{           27 }&\textbf{        23 
}\\ ham3\_102                   &   3 &              13 &                 11 &           29 &        26 &\textbf{            26 }&\textbf{         23 }&\textbf{           26 }&\textbf{        23 
}\\ mod5d1\_63                  &   5 &              13 &                 13 &\textbf{           33 }&\textbf{        31 }&            34 &\textbf{         31 }&\textbf{           33 }&\textbf{        31 
}\\
\hline
\end{longtable}

\begin{longtable}{lc|rr||rr|rr|rr|rr|rr}
\caption{Initial placement comparison of QUBO to the methods available in Qiskit, using IBM Melbourne. Reported depths take into account all gates (not just CX). We bold the best performing method in each row (for ties, each tieing method is bolded).}
\label{tab:qiskit_table} \\
\multicolumn{2}{c}{} & \multicolumn{2}{c}{\textbf{Initial}} & \multicolumn{2}{c}{\textbf{Trivial}} & \multicolumn{2}{c}{\textbf{Dense}} & \multicolumn{2}{c}{\textbf{Noise}} & \multicolumn{2}{c}{\textbf{SABRE}} & \multicolumn{2}{c}{\textbf{QUBO}}\\
 Circuit              &   n &  Depth &  CX count &  Depth &  CX &  Depth &  CX &  Depth &  CX &  Depth &  CX &  Depth &  CX \\
\toprule
\endfirsthead

\multicolumn{2}{c}{} & \multicolumn{2}{c}{\textbf{Initial}} & \multicolumn{2}{c}{\textbf{Trivial}} & \multicolumn{2}{c}{\textbf{Dense}} & \multicolumn{2}{c}{\textbf{Noise}} & \multicolumn{2}{c}{\textbf{SABRE}} & \multicolumn{2}{c}{\textbf{QUBO}}\\
 Circuit              &   n &  Depth &  CX count &  Depth &  CX &  Depth &  CX &  Depth &  CX &  Depth &  CX &  Depth &  CX \\
\toprule
\endhead
ex1\_226              &   6 &               5 &                  5 &              30 &           47 &            23 &         32 &            12 &        17 &\textbf{             8 }&\textbf{         11 }&\textbf{            8 }&\textbf{        11 
}\\ graycode6\_47         &   6 &               5 &                  5 &\textbf{               5 }&\textbf{            5 }&            12 &         14 &            13 &         20 &\textbf{             5 }&         11 &\textbf{            5 }&\textbf{         5 
}\\ xor5\_254             &   6 &               5 &                  5 &              30 &           47 &            23 &         32 &            12 &         17 &\textbf{             8 }&\textbf{         11 }&\textbf{            8 }&\textbf{        11 
}\\ ex-1\_166             &   3 &              12 &                  9 &              29 &\textbf{           24 }&\textbf{            27 }&\textbf{         24 }&            28 &\textbf{         24 }&            29 &\textbf{         24 }&\textbf{           27 }&\textbf{        24 
}\\ 4mod5-v0\_20          &   5 &              12 &                 10 &              39 &           43 &            27 &         25 &            30 &         31 &\textbf{            21 }&         22 &\textbf{           21 }&\textbf{        19 
}\\ 4mod5-v1\_22          &   5 &              12 &                 11 &              36 &           35 &\textbf{            21 }&\textbf{         20 }&            33 &         32 &            24 &         26 &\textbf{           21 }&\textbf{        20 
}\\ ham3\_102             &   3 &              13 &                 11 &              27 &           23 &            25 &         23 &            27 &         23 &\textbf{            22 }&\textbf{         20 }&           23 &\textbf{        20 
}\\
\hline
\end{longtable}


\end{document}